\documentclass[pre,aps,twocolumn,showpacs,superscriptaddress,floatfix]{revtex4-2}

\usepackage{mathtools}
\usepackage[usenames,dvipsnames]{xcolor}
\usepackage{amsmath}
\usepackage{amssymb,mathrsfs}
\usepackage{graphicx}
\usepackage[colorlinks=true]{hyperref}
\usepackage{soul}
\usepackage{xcolor}

\usepackage{float}

\newcommand{\prt}{\partial}

\newcommand{\om}{\omega}

\DeclareMathOperator{\arccosh}{arccosh}

\begin{document}

\title{Asymptotic integrability and Hamilton theory of soliton's motion along large-scale
background waves}

\author{A.~M.~Kamchatnov}
\affiliation{Institute of Spectroscopy, Russian Academy of Sciences, Troitsk, Moscow, 108840, Russia}
\affiliation{Skolkovo Institute of Science and Technology, Skolkovo, Moscow, 143026, Russia}

\begin{abstract}
We consider the problem of soliton-mean field interaction for the class of asymptotically
integrable equations, where the notion of the asymptotic integrability means that the Hamilton
equations for the high-frequency wave packet's propagation along a large-scale
background wave have an integral of motion. Using the Stokes remark, we transform this
integral to the integral for the soliton's equations of motion and then derive the Hamilton
equations for the soliton's dynamics in a universal form expressed in terms of the Riemann
invariants for the hydrodynamic background wave. The physical properties are
specified by the concrete expressions for the Riemann invariants. The theory is illustrated
by its application to the soliton's dynamics which is described by the Kaup-Boussinesq system.
\end{abstract}



\maketitle

\section{Introduction}

As is known, the discovery of the inverse scattering transform method \cite{ggkm-67,lax-68,zs-71}
allowed one to distinguish an important class of completely integrable wave equations
\cite{zf-71,gardner-71} with remarkable mathematical properties (see, e.g., \cite{dickey,ft-07}
and references therein). The special properties of these equations
found a number of applications to problems of solitons' dynamics in different physical
situations as, e.g., interaction of solitons, kinetic theory of soliton gases, evolution of
soliton's parameters under the action of small perturbations. It is remarkable that the property
of complete integrability is preserved by some approximate procedures applied to the
completely integrable equations. For example, asymptotic perturbation theory applied to the
completely integrable soliton equations yields again completely integrable equations \cite{zk-86}
and averaging of periodic solutions leads to the Whitham modulation equations which also have
the property of complete integrability \cite{dn-93}. It is natural to expect
that other approximation schemes can also reduce a complicated problem described by a
completely integrable system to a much simpler approximate system of completely integrable
Hamilton equations.

In this paper, we shall consider the problem of propagation of a narrow soliton along a wide
and smooth large-scale background wave. In this case, the complete integrability is understood
in the restricted sense called `asymptotic integrability' introduced in Ref.~\cite{kamch-24}. It means
that the Hamilton equations, that describe propagation of short-wavelength packets along a smooth
large-scale background, have an integral of motion, so that existence of such an integral allows
one to integrate these equations in a closed form \cite{sk-23}. Using the Stokes remark
\cite{stokes} that the soliton's tails and linear harmonic waves obey the same equations,
the above mentioned integral can be cast to the relationship between the soliton's velocity and its
inverse half-width. This gives a convenient method of solving problems of interaction of
solitons with a background wave (see, e.g., Ref.~\cite{acehl-23,ks-23a}). A different approach
based on the Whitham modulation theory was developed in Refs.~\cite{she-18,seh-21}. As was shown
in Refs.~\cite{kamch-24b,ks-24,kamch-24c}, the above mentioned relationship allows one
to obtain the Hamilton equations for motion of solitons of the nonlinear Schr\"{o}dinger (NLS),
derivative nonlinear Schr\"{o}dinger (DNLS), and ``magnetic soliton'' equations. We will show here
that this approach
can be generalized in such a way that the Hamilton equations are obtained for general classes of
equations that satisfy the conditions of asymptotic integrability. These general equations are
expressed in terms of the Riemann invariants for the dispersionless (hydrodynamic) equations governing
the large-scale evolution. The concrete physical realizations of this scheme depend on the
expressions for the Riemann invariants in terms of the physical variables of the problem.

In the next section \ref{as-int} we will discuss the concept of asymptotic integrability and in
section \ref{sol-mot} we will apply this method to the problem of soliton's dynamics in the
general formulation applicable to any asymptotically integrable equation. This approach is
is illustrated in section \ref{example} of soliton's motion along a large scale wave described
by the Kaup-Boussinesq system.

\section{Asymptotic integrability}\label{as-int}

Let the physical system under consideration be described by two wave variables $\rho$ (`density')
and $u$ (`flow velocity'). We suppose that dynamics of this system obeys the equations which
can be written in the form
\begin{equation}\label{eq1}
\begin{split}
  &\rho_t+F(\rho,u,\rho_x,u_x,\ldots)=0,\\
  &u_t+G(\rho,u,\rho_x,u_x,\ldots)=0,
  \end{split}
\end{equation}
where dots denote higher order $x$-derivatives. We assume that this system has a stationary and
uniform solution $\rho=\rho_0=\mathrm{const}$, $u=\rho_0=\mathrm{const}$, where $\rho_0,u_0$ belong
to some intervals (usually $\rho_0>0$), so we have a continuous set of such solutions. Small
deviations $\rho'=\rho-\rho_0, u'=u-u_0$ from these stationary and uniform states obey
Eqs.~(\ref{eq1}) linearized with respect to $\rho',u'$, and then the linear harmonic waves
\begin{equation}\label{eq1a}
  \rho',u'\propto\exp[i(kx-\om t)]
\end{equation}
can propagate along these states, if the dispersion relation
\begin{equation}\label{eq2}
  \om=\om_{\pm}(k,\rho_0,u_0)
\end{equation}
is fulfilled. This relation is an important physical characteristic of the system (\ref{eq1}) and
it describes its dispersive properties.

Another important limit is provided by the so-called dispersionless (hydrodynamic) approximation
when one can neglect all the higher-order derivatives or powers of derivatives in Eqs.~(\ref{eq1}).
We assume that the system arising in this limit can be written in a quasi-linear form
\begin{equation}\label{eq3}
  \left(
          \begin{array}{c}
            \rho \\
            u \\
          \end{array}
        \right)_t+ \mathbb{A}
        \left(
          \begin{array}{c}
            \rho \\
            u \\
          \end{array}
        \right)_x=0,\qquad
        \mathbb{A}=\left(
                     \begin{array}{cc}
                       a_{11} & a_{12} \\
                       a_{21} & a_{22} \\
                     \end{array}
                   \right),
\end{equation}
where $a_{ij}=a_{ij}(\rho,u)$. Such a system can be transformed to a diagonal
Riemann form~\cite{ry-83}
\begin{equation}\label{eq4}
  \frac{\prt r_+}{\prt t}+v_+\frac{\prt r_+}{\prt x}=0,\qquad
  \frac{\prt r_-}{\prt t}+v_-\frac{\prt r_-}{\prt x}=0
\end{equation}
for so-called Riemann invariants $r_{\pm}=r_{\pm}(\rho,u)$. The physical variables $\rho,u$
can be expressed as functions of $r_{\pm}$, $\rho=\rho(r_+,r_-),u=u(r_+,r_-)$, and their
substitution into Eq.~(\ref{eq2}) gives the dispersive relation in the form
\begin{equation}\label{eq5}
  \om=\om_{\pm}(k,r_+,r_-).
\end{equation}
The characteristic velocities $v_{\pm}$ in Eqs.~(\ref{eq4}) are the long wavelength limits
of the phase velocities of the harmonic waves, so we have
\begin{equation}\label{eq6}
  v_{\pm}=\lim_{k\to0}\frac{\om_{\pm}(k,r_+,r_-)}{k}.
\end{equation}

Now we turn to the definition of the asymptotic integrability. Let us assume that we know
some solution of Eqs.~(\ref{eq4}), which describes a large-scale hydrodynamic flow with a
characteristic size $l$, and we are interested in propagation of a short wavelength wave
packet along such a flow, provided the size $\Delta$ of the packet and the carrier wave's
wavelength $\sim2\pi/k$ satisfy the conditions
\begin{equation}\label{eq7}
  l\gg\Delta\gg2\pi/k.
\end{equation}
In this case, we can define a coordinate $x(t)$ of the packet with the accuracy $\Delta$ and
the wave number $k$ with the accuracy $\sim2\pi/\Delta\ll k$, so that with this accuracy the
packet's dynamics obeys the Hamilton equations \cite{LL-2,ko-90}
\begin{equation}\label{eq8}
  \frac{dx}{dt}=\frac{\prt\om}{\prt k}, \qquad \frac{dk}{dt}=-\frac{\prt\om}{\prt x},
\end{equation}
where $\om$ depends on $x$ and $t$ via the functions $r_{\pm}=r_{\pm}(x,t)$ and $x$ denotes
here the coordinate of the packet. The first Eq.~(\ref{eq8}) corresponds to the well-known
definition of the group velocity
\begin{equation}\label{eq9}
  v_g=\frac{\prt\om}{\prt k},
\end{equation}
and the second Eq.~(\ref{eq8}) describes refraction, i.e., change of the wavelength in a
non-uniform medium. The systems (\ref{eq4}) and (\ref{eq8}) determine propagation of a
packet along a non-uniform and time-dependent background wave for given initial data
$r^{(0)}_{\pm}=r_{\pm}(x,0)$, $x_0=x(0),k_0=k(0)$. So far we have not imposed any conditions
on the dispersion relation (\ref{eq5}) in this general formulation of the problem.

Now we introduce into the theory an additional condition which means integrability of the
Hamilton equations (\ref{eq8}). Namely, we suppose that there exists an integral $k=k(r_+,r_-)$
of these equations which does not depend on a particular choice of the background wave
solution $r_{\pm}=r_{\pm}(x,t)$ of Eqs.~(\ref{eq4}). To express this condition in analytical
form, let us consider two moments of time separated by a small interval $dt$ and,
consequently, by a small distance $dx=v_gdt$ between the corresponding locations of the
packet. Then the differences of the values of the Riemann invariants at the locations of
the packet at these two moments of time are equal to
$$
dr_{\pm}=\frac{\prt r_{\pm}}{\prt x}dx+\frac{\prt r_{\pm}}{\prt t}dt=
\frac{\prt r_{\pm}}{\prt x}(v_g-v_{\pm})dt,
$$
where we used the hydrodynamic equations (\ref{eq4}). Hence, we find
$$
\frac{\prt r_{\pm}}{\prt x}dt=-\frac{dr_{\pm}}{v_{\pm}-v_g}.
$$
Consequently, the second Hamilton equation (\ref{eq8}) yields the following
expression for the change of the wave number along the packet's path
\begin{equation}\nonumber
  \begin{split}
  dk=-\frac{\prt\om}{\prt x}dt&=-\left(\frac{\prt\om}{\prt r_+}\frac{\prt r_+}{\prt x}+
  \frac{\prt\om}{\prt r_-}\frac{\prt r_-}{\prt x}\right)dt\\
  &=\frac{\prt\om/\prt r_+}{v_+-v_g}dr_++\frac{\prt\om/\prt r_-}{v_--v_g}dr_-.
  \end{split}
\end{equation}
On the other hand, our supposition that the integral $k=k(r_+,r_-)$ does exist gives
$$
dk=\frac{\prt k}{\prt r_+}dr_+ +\frac{\prt k}{\prt r_-}dr_-.
$$
Since we consider an arbitrary solution of Eqs.~(\ref{eq4}), the differentials $dr_+$ and
$dr_-$ are independent of each other and comparison of two expressions for $dk$ yields
the system of equations
\begin{equation}\label{eq10}
  \frac{\prt k}{\prt r_+}=\frac{\prt\om/\prt r_+}{v_+-v_g},\quad
  \frac{\prt k}{\prt r_-}=\frac{\prt\om/\prt r_-}{v_--v_g},
\end{equation}
which determines the function $k=k(r_+,r_-)$, i.e., the integral of Eqs.~(\ref{eq8}).
Taking into account Eqs.~(\ref{eq6}) and (\ref{eq9}), we see that this integral is
determined by the dispersion relation (\ref{eq5}). It is clear that the solution of
Eqs.~(\ref{eq10}) exists if the derivatives of $k$ with respect to $r_+$ and $r_-$
commute,
\begin{equation}\label{eq11}
  \frac{\prt}{\prt r_-}\left(\frac{\prt k}{\prt r_+}\right)=
   \frac{\prt}{\prt r_+}\left(\frac{\prt k}{\prt r_-}\right).
\end{equation}
We call this the asymptotic integrability condition, since it is based on the asymptotic
theory of propagation of short wavelength packets along a non-uniform background wave.
The condition (\ref{eq11}) imposes heavy restrictions on the form of the dispersion relation.
We confine ourselves by two examples of the dispersion relations that satisfy this condition.

In the first example the dispersion relation reads
\begin{equation}\label{eq12}
  \om=k\left(r_++r_-\pm\frac12\sqrt{(r_+-r_-)^2+\sigma{k^2}}\right),
\end{equation}
where $\sigma=\pm1$ determines the sigh of the dispersion.
Then, assuming $r_+>r_-$, we find the characteristic velocities
\begin{equation}\label{eq13}
  v_{\pm}=r_++r_-\pm\frac12(r_+-r_-),
\end{equation}
and the solution of Eqs.~(\ref{eq10}) has the form
\begin{equation}\label{eq14}
  \sigma k^2=4(q-r_+)(q-r_-),
\end{equation}
where $q$ is an integration constant.

In the second example we have
\begin{equation}\label{eq15}
\begin{split}
  \om=&k\Bigg\{r_++r_-+\frac{\sigma k^2}8\\
  &\pm\sqrt{\left(\frac{r_++r_-}2+\frac{\sigma k^2}8\right)^2-r_+r_-}\Bigg\},
  \end{split}
\end{equation}
so that for $r_+>r_-$ the characteristic velocities are given again by Eqs.~(\ref{eq13})
and the integral has the form
\begin{equation}\label{eq16}
 \sigma k^2=\frac{4}{q}(q-r_+)(q-r_-),
\end{equation}
where $q$ is an integration constant and again $\sigma=\pm1$.

As was shown in Refs.~\cite{kamch-24,ks-24b}, these integrals are related with the
quasi-classical limit of Lax pairs of completely integrable equations in the
Ablowitz-Kaup-Newell-Segur (AKNS) scheme \cite{akns-74}. If we write an AKNS Lax pair
in the scalar form (see, e.g., \cite{kk-02})
\begin{equation}\label{eq17}
  \phi_{xx}=\mathcal{A}\phi,\qquad \phi_t=-\frac12\mathcal{B}_x\phi+\mathcal{B}\phi_x,
\end{equation}
and assume
\begin{equation}\label{eq18}
  \begin{split}
  &\mathcal{A}=-\frac14k^2(r_+,r_-,q),\\
  &\mathcal{B}=-\frac{\om(k(r_+,r_-,q),r_+,r_-)}{k(r_+,r_-,q)},
  \end{split}
\end{equation}
where $k^2$ is defined by either Eq.~(\ref{eq14}) or Eq.~(\ref{eq16}), then the
compatibility condition $(\phi_{xx})_t=(\phi_t)_{xx}$ with $q$ playing a role of
the spectral parameter yields the completely integrable system of equations.
In case of Eq.~(\ref{eq14})  with
\begin{equation}\label{eq19}
  \mathcal{A}=-\sigma(q-r_+)(q-r_-),\quad \mathcal{B}=-q-\frac12(r_++r_-)
\end{equation}
we get the system
\begin{equation}\label{eq20}
  \begin{split}
   (r_++r_-)_t&+\frac32(r_++r_-)(r_++r_-)_x
      -(r_+r_-)_x=0,\\
   (r_+r_-)_t&+r_+r_-(r_++r_-)_x
   +\frac12(r_++r_-)(r_+r_-)_x\\ &   +\frac{\sigma}{4}(r_++r_-)_{xxx}=0
  \end{split}
\end{equation}
corresponding to the dispersion relation (\ref{eq12}), and in case of (\ref{eq16}) with
\begin{equation}\label{eq21}
  \mathcal{A}=-\frac{\sigma}{q}(q-r_+)(q-r_-),\quad \mathcal{B}=-q-\frac12(r_++r_-)
\end{equation}
we obtain
\begin{equation}\label{eq22}
  \begin{split}
   (r_++r_-)_t&+\frac32(r_++r_-)(r_++r_-)_x\\
   &-(r_+r_-)_x-\frac{\sigma}{4}(r_++r_-)_{xxx}=0,\\
   (r_+r_-)_t&+r_+r_-(r_++r_-)_x\\
   &+\frac12(r_++r_-)(r_+r_-)_x=0,
  \end{split}
\end{equation}
and this system corresponds to the dispersion relation (\ref{eq15}) for the linear
harmonic waves.

These completely integrable systems acquire physical meaning after relating the
Riemann invariants to the physical variables. For example, if we take
\begin{equation}\label{eq23}
  r_+=\frac{u}{2}+\sqrt{\rho},\qquad r_-=\frac{u}{2}-\sqrt{\rho},
\end{equation}
the we get from Eqs.~(\ref{eq20}) the system
\begin{equation}\label{eq24}
  \rho_t+(\rho u)_x-\frac{\sigma}4u_{xxx}=0,\qquad u_t+uu_x+\rho_x=0.
\end{equation}
This is the Kaup-Boussinesq system \cite{kaup-75} which describes the shallow water
dynamics, so that the linear waves propagate according to the dispersion relation
\begin{equation}\label{eq25}
  \om=k\left(u\pm\sqrt{\rho+\frac{\sigma k^2}{4}}\right).
\end{equation}
Substitution of the same expressions (\ref{eq23}) for the Riemann invariants into
Eqs.~(\ref{eq22}) yields another completely integrable system
\begin{equation}\label{eq26}
  \begin{split}
  & \rho_t+(\rho u)_x-\frac{\sigma}8uu_{xxx}=0,\\
  & u_t+uu_x+\rho_x-\frac{\sigma}4u_{xxx}=0,
  \end{split}
\end{equation}
which corresponds to the dispersion relation
\begin{equation}\label{eq27}
  \om=k\left(u+\frac{\sigma k^2}{8}\pm\sqrt{\rho+\frac{k^2}{8}\left(u+\frac{\sigma k^2}{8}\right)}\right).
\end{equation}
We are not aware if the system (\ref{eq26}) appeared in the physical literature earlier.

Another definition of the Riemann invariants,
\begin{equation}\label{eq28}
  r_+=\frac{u}{2}+\sqrt{\frac{u^2}{4}-\rho},\quad r_-=\frac{u}{2}-\sqrt{\frac{u^2}{4}-\rho},
\end{equation}
yields after substitution into Eqs.~(\ref{eq20}) the system
\begin{equation}\label{eq29}
  \begin{split}
   &u_t+\frac32uu_x-\rho_x=0,\\
   &\rho_t+\rho u_x+\frac12u\rho_x+\frac{\sigma}4u_{xxx}=0
  \end{split}
\end{equation}
corresponding to the dispersion relation
\begin{equation}\label{eq30}
  \om=k\left(u+\sqrt{\frac{u^2}{4}-\rho+\frac{\sigma k^2}{4}}\right)
\end{equation}
for linear waves. Substitution of Eqs.~(\ref{eq28}) into Eqs.~(\ref{eq22}) reproduces the
Zakharov-Ito system \cite{zakh-80,ito-82}
\begin{equation}\label{31}
  \begin{split}
   &u_t+\frac32uu_x-\rho_x-\frac{\sigma}4u_{xxx}=0,\\
   &\rho_t+\rho u_x+\frac12u\rho_x=0,
  \end{split}
\end{equation}
corresponding to the dispersion relation
\begin{equation}\label{eq32}
  \om=k\left(u+\frac{\sigma k^2}{8}\pm\sqrt{\left(\frac{u}{2}+\frac{\sigma k^2}{8}\right)^2-\rho}\right).
\end{equation}

As one more example, let us take the Riemann invariants in the form
\begin{equation}\label{eq33}
  r_{\pm}=vw\pm\sqrt{(1-v^2)(1-w^2)}.
\end{equation}
They appear after diagonalization of the Ovsyannikov equations \cite{ovs-79}
\begin{equation}\label{eq34}
  v_t-[w(1-v^2)]_x=0,\qquad w_t-[v(1-w^2)]_x=0,
\end{equation}
which describe dispersionless dynamics of long waves on the surface separating two thin
layers of liquids with very small difference between their densities. Here $w$ corresponds to
deviation of the separation surface from the horizontal plane and $v$ denotes the relative
flow velocity in the two layers. Substitution of Eqs.~(\ref{eq33}) into Eqs.~(\ref{eq20})
yields the system
\begin{equation}\label{eq35}
  \begin{split}
   &v_t-[w(1-v^2)]_x+\frac{\sigma v(vw)_{xxx}}{4(v^2-w^2)}=0,\\
   &w_t-[v(1-w^2)]_x-\frac{\sigma w(vw)_{xxx}}{4(v^2-w^2)}=0,
  \end{split}
\end{equation}
which corresponds to the dispersive relation
\begin{equation}\label{eq36}
  \om=k\left(2vw\pm\sqrt{(1-v^2)(1-w^2)+\frac{\sigma k^2}{4}}\right)
\end{equation}
for linear harmonic waves. Substitution of Eqs.~(\ref{eq33}) into Eqs.~(\ref{eq22})
gives the system
\begin{equation}\label{eq37}
  \begin{split}
  & v_t-[w(1-v^2)]_x+\frac{\sigma w(vw)_{xxx}}{4(v^2-w^2)}=0,\\
   &w_t-[v(1-w^2)]_x-\frac{\sigma v(vw)_{xxx}}{4(v^2-w^2)}=0
  \end{split}
\end{equation}
with another dispersion relation
\begin{equation}\label{eq38}
  \om=k\left(2vw\pm\sqrt{(1-v^2)(1-w^2)+\frac{\sigma k^2}{4}\left(vw+\frac{\sigma k^2}{16}\right)}\right).
\end{equation}
Thus, we have obtained two completely integrable dispersive generalizations of the
Ovsyannikov equations. Another approach to completely integrable systems with the
general spectral problem in the form of the first Eq.~(\ref{eq17}) was developed
in the theory of so-called `energy dependent potentials' (see, e.g., Ref.~\cite{pavlov-14}
and references therein).

Lax pairs (\ref{eq19}) and (\ref{eq21}) do not depend on the $x$-derivatives of the Riemann
invariants. For many completely integrable equations in the AKNS scheme, the Lax pairs do
depend on the $x$-derivatives and then expressions (\ref{eq19}) and (\ref{eq21}) represent
just quasiclassical limits of the exact formulas for $\mathcal{A}$ and $\mathcal{B}$
\cite{kamch-24,ks-24b}. As a rule, soliton solutions of these integrable equations are studied
separately for every particular case. However, the general formulas (\ref{eq19}) and
(\ref{eq21}) for the quasiclassical limit of Lax pairs suggests that some properties of the
soliton solutions can also be expressed in the general form. For example, we have
studied in Refs.~\cite{kamch-24b,ks-24,kamch-24c} the Hamiltonian theory of soliton's
motion for NLS, DNLS, and easy-plane Landau-Lifshitz equations and found that the
expressions for the canonical momentum and Hamiltonian have evident similarities.
Therefore, it seems very plausible that they can be expressed in a very general form in
terms of the Riemann invariants for the dispersionless limit of equations under
consideration.  We will obtain these expressions in the next section \ref{sol-mot} of the paper.

\section{Hamilton theory of soliton's motion}\label{sol-mot}

Let the soliton solution correspond to its propagation with velocity $V$. Then far
enough from the soliton's center the physical variables $\rho,u$ approach to their
background values at the
soliton's tails exponentially with increase of $|x-Vt|$,
\begin{equation}\label{eq40}
  \rho-\rho_0,u-u_0\propto\exp[-\kappa|x-Vt|],
\end{equation}
where the parameter $\kappa$ is called the soliton's `inverse half-width'. As was
noticed by Stokes \cite{stokes}, the small-amplitude deviations $ \rho-\rho_0,u-u_0$
from the background state obey the same linearized equations as the linear harmonic
wave (\ref{eq1a}) and, consequently, the soliton velocity $V$ can be expressed via
the dispersion relation (\ref{eq5}): comparison of Eqs.~(\ref{eq1a}) and (\ref{eq40})
yields
\begin{equation}\label{eq41}
  V=\frac{\om(i\kappa,r_+,r_-)}{i\kappa},
\end{equation}
where it is assumed that $r_+,r_-$ change little across the soliton at distances about
$\kappa^{-1}\ll l$.

The condition of asymptotic integrability has led us to two integrals (\ref{eq14}) and
(\ref{eq16}) which are solutions of Eqs.~(\ref{eq10}). From a formal point of view,
the variable $k$ in these equations can be complex, so, following Stokes' consideration,
we can continue these solutions to the region $k\to i\kappa$ and obtain the expressions
\begin{equation}\label{eq42}
  \sigma\kappa^2=-4(q-r_+)(q-r_-)
\end{equation}
and
\begin{equation}\label{eq43}
  \sigma\kappa^2=-\frac{4}{q}(q-r_+)(q-r_-)
\end{equation}
for the soliton's inverse half-width $\kappa$ as a function of the background Riemann
invariants. It is evident that these relations are correct with good enough accuracy as long as the
soliton's width $\sim\kappa^{-1}$ is much smaller than the characteristic length $l$ at
which the background large-scale wave considerably changes. Substitution of Eqs.~(\ref{eq42})
and (\ref{eq43}) into Eq.~(\ref{eq41}) gives in both cases the same simple formula
($q>(r_++r_-)/2$)
\begin{equation}\label{eq44}
  V=\frac{dx}{dt}=\frac{\om(i\kappa,r_+,r_-)}{i\kappa}=q+\frac12(r_++r_-).
\end{equation}
This formula allows one to find the soliton's path $x=x(t)$ for known solutions
$r_+=r_+(x,t),r_-=r_-(x,t)$ of the dispersionless equations (\ref{eq4})
(see discussions of some particular cases in
Refs.~\cite{ks-24,kamch-24c,she-18,seh-21,ik-22}).

For taking into account the action of external potentials on the background wave flow,
we have to develop first the Hamiltonian theory of the soliton's dynamics. To this end,
it is convenient to consider the cases (\ref{eq42}) and (\ref{eq43}) separately. At first,
we will consider the case with $\sigma=+1$ and rewrite the identity (\ref{eq42}) in the form
\begin{equation}\label{eq45}
  \left[\frac{2q-(r_++r_-)}{r_+-r_-}\right]^2+\left(\frac{\kappa}{r_+-r_-}\right)^2=1.
\end{equation}
It is convenient to define a new variable $\phi$ so that
\begin{equation}\label{eq46}
  \begin{split}
  & q=\frac12(r_++r_-)+\frac12(r_+-r_-)\cos\phi,\\
  & \kappa=(r_+-r_-)\sin\phi.
  \end{split}
\end{equation}
Now we interpret Eq.~(\ref{eq44}) as the Hamilton equation
\begin{equation}\label{eq47}
  \frac{dx}{dt}=r_++r_-+\frac12(r_+-r_-)\cos\phi=\frac{\prt H}{\prt p},
\end{equation}
where only $\phi$ depends on the soliton's canonical momentum $p$. We will look 
for the expression for the canonical momentum $p$ in the form
\begin{equation}\label{eq48}
  p=(r_+-r_-)^2f(\phi).
\end{equation}
Then integration of Eq.~(\ref{eq47}) yields
\begin{equation}\label{eq49}
  H=(r_++r_-)p+\frac12(r_+-r_-)^3\int\cos\phi f'(\phi)d\phi.
\end{equation}

The function $f(\phi)$ can be found from the second Hamilton equation
\begin{equation}\label{eq50}
  \frac{dp}{dt}=-\frac{\prt H}{\prt x}.
\end{equation}
To simplify the arising equation, we obtain first two other useful formulas.
Differentiation of the first identity (\ref{eq46}) along the soliton's path gives
\begin{equation}\nonumber
  \begin{split}
  & (r_++r_-)_t+\left[r_++r_-+\frac12(r_+-r_-)\cos\phi\right](r_++r_-)_x\\
  &+\Bigg\{(r_+-r_-)_t+\left[r_++r_-+\frac12(r_+-r_-)\cos\phi\right]\\
  &\times(r_+-r_-)_x\Bigg\}\cos\phi
  -(r_+-r_-)\sin\phi\frac{d\phi}{dt}=0.
  \end{split}
\end{equation}
We eliminate $r_{+,t}$ and $r_{-,t}$ with the use of Eqs.~(\ref{eq4}) and (\ref{eq13}),
so after evident simplifications we obtain
\begin{equation}\label{eq51}
  \frac{d\phi}{dt}=-\frac12(r_+-r_-)_x\sin\phi.
\end{equation}
Besides that, differentiation of Eq.~(\ref{eq48}) with respect to $x$ for
$p=\mathrm{const}$ gives
\begin{equation}\label{eq52}
  \left(\frac{\prt\phi}{\prt x}\right)_p=-\frac{2(r_+-r_-)_x}{r_+-r_-}\frac{f}{f'}.
\end{equation}
Now, after substitution of Eqs.~(\ref{eq48}) and (\ref{eq49}) into Eq.~(\ref{eq50}),
we obtain after simple transformations with account of Eqs.~(\ref{eq51}) and (\ref{eq52})
the following equation
\begin{equation}\label{eq53}
  f'(\phi)\sin\phi=3\int^{\phi}\cos\phi\,f'(\phi)d\phi,
\end{equation}
which can be readily solved to give
\begin{equation}\label{eq54}
  f(\phi)=C(\phi-\sin\phi\cos\phi),
\end{equation}
where $C$ is an integration constant (another integration constant can be included
into the definition of $\phi$). For convenience of comparison with the previous
results, we choose $C=1/2$ and obtain
\begin{equation}\label{eq55}
  p=\frac12(r_+-r_-)^2(\phi-\sin\phi\cos\phi),
\end{equation}
\begin{equation}\label{eq56}
  H=(r_++r_-)p+\frac16(r_+-r_-)^3\sin^3\phi,
\end{equation}
where $\phi$ is related to the soliton's velocity $V=\dot{x}$ by the formula
\begin{equation}\label{eq57}
  \phi=\arccos\frac{2(\dot{x}-r_+-r_-)}{r_+-r_-}
\end{equation}
following from Eq.~(\ref{eq47}). Thus, the canonical momentum and the Hamiltonian
are defined in a parametric form. If the solution of the Hamilton equations is found,
then the dependence $\kappa=\kappa(t)$ of the soliton's inverse half-width on time
along the soliton's path is given by the second formula (\ref{eq46}). In many
concrete situations, $\kappa$ is related to the soliton's amplitude which, consequently,
can also be considered as known.

The theory for the solution (\ref{eq43}) of the asymptotical integrability equations
is very similar: we rewrite this identity with $\sigma=+1$ in the form
\begin{equation}\label{eq58}
  \left[\frac{2q-(r_++r_-)}{r_+-r_-}\right]^2+
  \left(\frac{\kappa\sqrt{q}}{r_+-r_-}\right)^2=1
\end{equation}
and introduce $\phi$ by the formulas
\begin{equation}\label{eq59}
  \begin{split}
  & q=\frac12(r_++r_-)+\frac12(r_+-r_-)\cos\phi,\\
  & \kappa=\frac{r_+-r_-}{\sqrt{q}}\sin\phi.
  \end{split}
\end{equation}
Since the first formula here coincides with the first formula in Eqs.~(\ref{eq46}),
the resulting expressions for the canonical momentum and the Hamiltonian coincide with
Eqs.~(\ref{eq55}) and (\ref{eq56}), correspondingly, but the expression for the
inverse half-width takes the form
\begin{equation}\label{eq60}
  \kappa=\frac{(r_+-r_-)\sin\phi}{\sqrt{r_+\cos^2(\phi/2)+r_-\sin^2(\phi/2)}}.
\end{equation}

In practice, if there are no external potentials, it is more convenient to solve
Eq.~(\ref{eq44}). For taking into account the external potentials, it is convenient to
transform the Hamilton equations to the Newton-like equation. Such a transformation
was done for a particular case of the NLS equation in Ref.~\cite{ik-22} and calculations
in the general case are actually the same, so the final result can be obtained from
Eq.~(21) of Ref.~\cite{ik-22} by the replacements $u\to r_++r_-$, $\rho\to\frac12(r_+-r_-)^2$.
Thus, we get the Newton equation in the form
\begin{equation}\label{eq61}
  \begin{split}
  2\ddot{x}&=\frac12(r_+-r_-)(r_+-r_-)_x+\dot{x}(r_++r_-)_x\\
  &+(r_++r_-)(r_++r_-)_x+2(r_{+,t}+r_{-,t})\\
  &+\frac{[(r_+-r_-)^2]_t+[(r_+-r_-)^2(r_++r_-)]_x}{2\sqrt{(r_+-r_-)^2-4(\dot{x}-r_+-r_-)^2}}\\
  &\times\arccos\frac{2(\dot{x}-r_+-r_-)}{r_+-r_-}.
  \end{split}
\end{equation}
As we see, in this approximation the soliton's dynamics is governed by the gradients
of the Riemann invariants. Of course, these gradients are considered as small enough not
to change the soliton's profile. If the external potentials are also small and do not
influence on the soliton's profile, then we can only take them into account in the
hydrodynamic equations which are generalized to the form
\begin{equation}\label{eq62}
\begin{split}
  \frac{\prt r_+}{\prt t}+\left[r_++r_-+\frac12(r_+-r_-)\right]\frac{\prt r_+}{\prt x}=-U_{+,x},\\
  \frac{\prt r_-}{\prt t}+\left[r_++r_--\frac12(r_+-r_-)\right]\frac{\prt r_-}{\prt x}=-U_{-,x},
  \end{split}
\end{equation}
where $U_+,U_-$ are to be obtained from the hydrodynamic equations for the physical variables.
Exclusion of the derivatives $r_{+,t},r_{-,t}$ from Eqs.~(\ref{eq61}) with the use of
Eq.~(\ref{eq62}) yields
\begin{equation}\label{eq63}
  \begin{split}
  2\ddot{x}&= (\dot{x}-r_+-r_-)(r_++r_-)_x\\
  &-\left[\frac14(r_+-r_-)^2+2(U_++U_-)\right]_x\\
  &-\frac{(r_+-r_-)](U_+-U_-)_x}{\sqrt{(r_+-r_-)^2-4(\dot{x}-r_+-r_-)^2}}\\
  &\times\arccos\frac{2(\dot{x}-r_+-r_-)}{r_+-r_-}.
  \end{split}
\end{equation}

This is the general Newton equation for description of soliton's dynamics for asymptotically
integrable equations with positive dispersion $\sigma=+1$. Its particular
cases are specified by the choice of the Riemann invariants in concrete physical situations.

In case of negative dispersion $\sigma=-1$ the calculations are very similar. Now we write 
the identity (\ref{eq42}) in the form
\begin{equation}\label{eq63b}
  \left[\frac{2q-(r_++r_-)}{r_+-r_-}\right]^2-\left(\frac{\kappa}{r_+-r_-}\right)^2=1
\end{equation}
and define a new variable $\phi$ in such a way that
\begin{equation}\label{eq63c}
  \begin{split}
  & q=\frac12(r_++r_-)+\frac12(r_+-r_-)\cosh\phi,\\
  & \kappa=(r_+-r_-)\sinh\phi.
  \end{split}
\end{equation}
Correspondingly, we obtain the expressions for the canonical momentum and the Hamiltonian 
\begin{equation}\label{eq55b}
  p=\frac12(r_+-r_-)^2(\sinh\phi\cosh\phi-\phi),
\end{equation}
\begin{equation}\label{eq56b}
  H=(r_++r_-)p+\frac16(r_+-r_-)^3\sinh^3\phi,
\end{equation}
where $\phi$ is related to the soliton's velocity $V=\dot{x}$ by the formula
\begin{equation}\label{eq57b}
  \phi=\arccosh\frac{2(\dot{x}-r_+-r_-)}{r_+-r_-}.
\end{equation}The Hamilton equations can be reduced to the Newton-like equation
\begin{equation}\label{eq61b}
  \begin{split}
  2\ddot{x}&=\frac12(r_+-r_-)(r_+-r_-)_x+\dot{x}(r_++r_-)_x\\
  &+(r_++r_-)(r_++r_-)_x+2(r_{+,t}+r_{-,t})\\
  &+\frac{[(r_+-r_-)^2]_t+[(r_+-r_-)^2(r_++r_-)]_x}{2\sqrt{4(\dot{x}-r_+-r_-)^2-(r_+-r_-)^2}}\\
  &\times\arccosh\frac{2(\dot{x}-r_+-r_-)}{r_+-r_-}.
  \end{split}
\end{equation}
Elimination of $r_{+,t}$ and $r_{-,t}$ with help of Eqs.~(\ref{eq62}) yields
\begin{equation}\label{eq63d}
  \begin{split}
  2\ddot{x}&= (\dot{x}-r_+-r_-)(r_++r_-)_x\\
  &-\left[\frac14(r_+-r_-)^2+2(U_++U_-)\right]_x\\
  &-\frac{(r_+-r_-)](U_+-U_-)_x}{\sqrt{4(\dot{x}-r_+-r_-)^2-(r_+-r_-)^2}}\\
  &\times\arccosh\frac{2(\dot{x}-r_+-r_-)}{r_+-r_-}.
  \end{split}
\end{equation}

\section{Example}\label{example}

Let us consider the Kaup-Boussinesq system (\ref{eq24}) ($\sigma=+1$) with addition of the 
external potential $U(x)$ which models the `force' acting in $x$-direction:
\begin{equation}\label{eq64}
  \rho_t+(\rho u)_x-\frac14u_{xxx}=0,\quad u_t+uu_x+\rho_x=-U_x.
\end{equation}
The equations for the hydrodynamic approximation with omitted term $-u_{xxx}/4$ after
transformation to the Riemann invariants (\ref{eq23}) take the form (\ref{eq62})
with $U_+=U_-=U/2$. Consequently, Eq.~(\ref{eq61}) simplifies to
\begin{equation}\label{eq66}
  2\ddot{x}=(\dot{x}-u)u_x-\rho_x-2U_x.
\end{equation}
Let the soliton propagate along the background with the fluid at rest ($u=0$), so that
the second Eq.~(\ref{eq64}) gives $\rho_x=-U_x$ and Eq.~(\ref{eq66}) reduces to
\begin{equation}\label{eq67}
  2\ddot{x}=\rho_x=-U_x.
\end{equation}
Its integration gives at once
\begin{equation}\label{eq68}
  \dot{x}^2=\rho(x)+\mathrm{const}=-U(x)+\mathrm{const}.
\end{equation}
On the other hand, the second Eq.~(\ref{eq46}) gives
\begin{equation}\label{eq69}
  \kappa^2=(r_+-r_-)^2-4(\dot{x}-r_+-r_-)^2=4(\rho-\dot{x}^2).
\end{equation}
Comparison with Eq.~(\ref{eq68}) shows that the soliton's inverse half-width $\kappa$
is constant along the soliton's path and
\begin{equation}\label{eq70}
  \dot{x}^2=\rho(x)-\frac{\kappa^2}{4}.
\end{equation}
Thus, the soliton's path is given implicitly by the formula
\begin{equation}\label{eq71}
  t-t_0=\int_{x_0}^x\frac{dx}{\sqrt{\rho(x)-\kappa^2/4}}
\end{equation}
for the soliton having the inverse half-width $\kappa$, when it propagates along the
profile $\rho(x)$ of the density. This profile is formed by the external
potential $U(x)$ according to Eq.~(\ref{eq68}).

The soliton solution of Eqs.~(\ref{eq24}) has the form of a dip $\rho(x-Vt)$ in the density
$\rho_0$ and a local distribution $u(x-Vt)$ of the flow velocity caused by the moving soliton.
This solution can be written in the form \cite{cikp-17}
\begin{equation}\label{eq73}
  \begin{split}
  &\rho(\xi)=\rho_0-2\mu(\xi)[\mu(\xi)-V],\\
  &u(\xi)=2(V-\mu(\xi)),
  \end{split}
\end{equation}
where
\begin{equation}\label{eq74}
  \mu(\xi)=V+\frac{\rho_0-V^2}{\sqrt{\rho_0}\,\cosh(\kappa\xi)+V}
\end{equation}
and
\begin{equation}\label{eq75}
  \xi=x-Vt,\qquad V=\sqrt{\rho_0-\kappa^2/4}.
\end{equation}
Consequently, we get the approximate solution for the soliton moving along a smooth
background $\rho=\rho(x)$ by means of replacements
\begin{equation}\label{eq76}
  \begin{split}
  &\rho_0\to\rho(x),\qquad V\to\sqrt{\rho(x)-\kappa^2/4},\\
  &\xi\to x-\int_{t_0}^{t}\sqrt{\rho(x(t))-\kappa^2/4}\,dt-x_0.
  \end{split}
\end{equation}
Although the soliton's inverse half-width remains constant, its amplitude depends on the
local density $\rho(x)$ as
\begin{equation}\label{eq77}
  a=2\left(\rho(x)-\sqrt{\rho(x)(\rho(x)-\kappa^2/4)}\right).
\end{equation}

In the small-amplitude limit $\kappa^2\ll\rho$, this solution reduces to the well-known
shallow dark KdV soliton. As was shown
in Ref.~\cite{ikcp-17}, the Kaup-Boussinesq system (\ref{eq24}) with $\sigma=+1$ can find 
applications to dynamics of solitons in two-component Bose-Einstein condensates.

\section{Conclusion}

The concept of complete integrability of nonlinear wave equations plays a crucial role in
soliton physics \cite{ggkm-67,lax-68,zs-71,zf-71,gardner-71,dickey,ft-07}. The notion of
asymptotic integrability restricts this concept to a particular case of propagation of
small-wavelength packets along a large-scale background wave. It turns out that this
notion leads directly to the quasi-classical limit of Lax pairs and, in the case of
systems with two wave variables, this limit admits quite a general description in terms
of Riemann invariants of the large-scale hydrodynamic approximation. Such a description
is only possible in exact form for special forms of the dispersion relation of linear waves,
and we presented here two such universal forms corresponding to a number of completely
integrable nonlinear wave equations. In all these cases, there exists an integral of motion
of the Hamilton equations that describe the packet's motion. Existence of such an integral
allows one to obtain the solution of these Hamilton equations in a closed form \cite{sk-23}.

According to Stokes' consideration  \cite{stokes}, the soliton's velocity can be expressed via
the linear dispersion relation as a function of the soliton's inverse half-width. In a similar
way, the mentioned above integral of motion for the packet's dynamics can be transformed to
the integral of motions for the dynamics of a soliton treated as a point-like particle
provided its width is much smaller than the characteristic scale of the background wave.
Due to knowledge of such an integral, one can obtain the Hamilton equations for the soliton's
dynamics and generalize them to situations when the background profiles are formed under
action of external potentials. We obtained the universal expressions for the canonical
momentum and Hamiltonian again in terms of the Riemann invariants for the background
dynamics. These universal expressions reproduce all particular cases studied earlier in
Refs.~\cite{kamch-24b,ks-24,kamch-24c}.

It is worth noticing that even if the condition of asymptotic integrability is not exactly
fulfilled, it may be satisfied approximately in the limit of large wave numbers of
the carrier wave. In this case, one can obtain an approximate integral of motion, which can be
used in applications to the packet's or soliton's dynamics. Besides that, it allows one to
formulate a generalized Bohr-Sommerfeld quantization rule that determines the asymptotic
velocities of solitons produced from an intensive initial pulse even for not completely
integrable equations \cite{kamch-23}. Thus, the concept of asymptotic integrability seems
quite fruitful, and one can hope that it will lead to many other interesting results.

\begin{acknowledgments}

I thank E.~A.~Kuznetsov and M.~V.~Pavlov for useful discussions.
This research is funded by the research project FFUU-2021-0003 of the Institute of Spectroscopy
of the Russian Academy of Sciences (Sections 1-2) and by the RSF grant number~19-72-30028
(Sections~3-4).

\end{acknowledgments}

\end{document}